\documentclass[aps,pra,reprint,superscriptaddress,nofootinbib,showpacs]{revtex4-2}

\usepackage{amsmath,amssymb,bm,mathtools}
\usepackage{physics}
\usepackage{braket}
\usepackage{graphicx}
\usepackage{booktabs}
\usepackage{hyperref}
\usepackage{xcolor}

\begin{document}

\title{Information flow control in hybrid quantum networks}

\author{Michael Siomau}
\email{msiomau@ra.ac.ae}
\affiliation{Rabdan Academy, Abu Dhabi, United Arab Emirates}
\affiliation{Institute for Theoretical Physics, ETH Zurich, Switzerland}

\date{\today}

\begin{abstract}
Quantum resources can change the dynamical features and operational behaviors of conventional networks. Long-distance entanglement supports information transfer between nodes lacking a direct connection. This entanglement link not only modifies the stationary effective connectivity of the network but can also be used to suppress undesirable classical dynamics. By examining information flow control mechanism, we demonstrate that classical collective behavior can be well-balanced by altering the initially distributed entanglement. We show the entanglement thresholds for two distinct control approaches: linear response control and Lyapunov functional controllability.
\end{abstract}


\maketitle

\section{Introduction}

Complex networks are currently viewed through two largely disjoint lenses. On one hand, classical network science \cite{Newman2010} typically characterizes the flow of information between nodes using entropic measures, such as transfer entropy, and relates it to the connectivity matrix and its spectral properties. Presenting collective dynamics, characterized by coupling terms that depend directly on the network structure, results in macroscopic phenomena, such as synchronization \cite{Arenas2008}. 

On the other hand, quantum networks are governed by quantum entanglement distribution, local operations and classical communication (LOCC), repeaters, quantum memory and quantum channel capacities \cite{Wehner2018}. The very notion of distance and connectivity becomes ambiguous once long-range entanglement links (e-links) are available \cite{Siomau2025}. Because these notions are at the core of the classical network science, there is no standard approach to study quantum networks \cite{Nokkala2024}. Original studies discuss entanglement percolation \cite{Acin2007}, the local Hamiltonian control \cite{Burgarth2009}, the quantum comb \cite{Chiribella2009} and the signaling power of quantum channels \cite{Santos2025}, to name a few. 

But, if we think of real applications \cite{Azuma2023}, a quantum network cannot be fully separated with its classical counterpart leading to the notion of hybrid quantum network, where classical topology and dynamics coexists with a reconfigurable layer of quantum entanglement. The mixed nature of these networks leads to interesting features, which cannot be seen in the associates. For instance, any hybrid network can be modified to become structurally controllable using only a polynomial number of LOCC-generated e-links, irrespective of the underlying classical topology \cite{Siomau2019}. This result raises a natural question that motivates the present work: is it possible to suppress unwanted classical dynamics by redistributing quantum resources?  

Control of hybrid quantum networks should be formulated at the interface between classical collective dynamics and quantum resource allocation. The present work develops such a formulation. The central idea is to describe a node's total control-relevant activity as the sum of three contributions: local classical influence arriving through physical edges, a network-wide classical collective term, and a quantum compensation term arising from LOCC-enabled redistribution of entanglement resources. Imposing approximate stationary on the total activity yields a controllability equation relating variations of classical collective behavior to variations of the quantum resource layer. We develop two solutions and exemplify them on a classical dense random graph subjected with distributed two-qubit entanglement. We explicitly show resource-scaling bounds for two control scenarios. 

\section{Hybrid Network}

Let us consider an undirected connected multigraph $G=(N, A, E)$ consisting of $N$ nodes, $A$ classical links and $E$ entanglement links. The classical connectivity of the graph is described with adjacency matrix $A\in\{0,1\}^{N\times N}$, where $A_{ij}=1$ when a physical channel exists between nodes $i$ and $j$ and $A_{ij}=0$ otherwise.

The quantum connectivity is represented by an effective entanglement graph, where an edge $(i,j)\in E$ indicates that nodes $i$ and $j$ share a usable entangled resource. In other words, it denotes a scalar entanglement resource associated with the pair $(i,j)$, which could be quantified with concurrence, negativity, distillable entanglement, entanglement of formation, or an entanglement-generation rate \cite{Horodecki2009}. We deliberately keep $E_{ij}$ abstract so that later considerations do not depend on one particular monotone.

In general, the entanglement graph is not restricted by classical connectivity, i.e. $E \not\subseteq A$, because long-range entanglement links can be generated over multi-hop physical paths by entanglement swapping and LOCC. The entanglement layer is therefore a dynamical object built on top of the physical substrate.

To model collective dynamics of the hybrid network, we use transfer entropy because it detects direct statistical dependence without requiring a pre-specified interaction model. If the transfer entropy between two classical nodes is denoted as $T_{j\to i}$, then information $I_i$ transferred to node $i$ through the classical channels is 
\begin{equation}
I_i=\sum_{j}A_{ij}T_{j\to i}.
\label{eq:localTE}
\end{equation}
A standard move in classical network science is to supplement Eq.~\eqref{eq:localTE} with a network-wide collective term that depends on the global structure of $A$ rather than on the immediate neighbors of $i$ alone. Three common choices are \cite{Newman2010}: (i) the simple neighborhood average $\alpha\sum_j A_{ij}/k_i$ with $k_i=\sum_j A_{ij}$ the degree of $i$; (ii) a Kuramoto-type phase-coupling term $\alpha\sum_j A_{ij}\sin(\theta_j-\theta_i)$; and (iii) a spectral term proportional to the principal eigenvector of $A$,
\begin{equation}
g_i(A) = \alpha\, v_1^{(i)}(A), \qquad A\, v_1 = \lambda_1\, v_1,
\label{eq:eigterm}
\end{equation}
with $\lambda_1$ the largest eigenvalue of $A$ (which exists and is non-degenerate for a connected network by the Perron--Frobenius theorem) and $v_1^{(i)}$ the $i$-th component of the associated eigenvector, i.e., the eigenvector centrality of node $i$. 

We adopt Eq.~\eqref{eq:eigterm} for the remainder of the paper because it is analytically tractable under structural perturbations of $A$, while (i) and (ii) can be recovered as limiting or averaged cases of the same construction. 

The total classical information activity is then
\begin{equation}
I_i^{\rm cl} = \sum_j A_{ij}\,T_{j\to i} + \alpha\, v_1^{(i)}(A).
\label{eq:classicaltotal}
\end{equation}

By analogy with Eq.~\eqref{eq:localTE} we can define quantum information $Q_i$ transferred to node $i$
\begin{equation}
Q_i=\sum_{k}E_{ik}\Lambda_{k\to i},
\label{eq:quantum}
\end{equation}
where $\Lambda_{j\to i}$ denote an effective LOCC-control coefficient describing the usable contribution of the entanglement resource between nodes $i$ and $j$. It is important to note that the quantum links do not inject classical signal flow; rather, they reshape the effective communication geometry by consuming, concentrating, and reallocating entanglement resources through LOCC. 

The total information flow that takes into account both classical and quantum contributions is thus
\begin{equation}
I_i^{\rm tot} = \sum_j A_{ij}\,T_{j\to i} + \alpha\, v_1^{(i)}(A) + \sum_k E_{ik}\Lambda_{k\to i} \, .
\label{eq:classicaltotal}
\end{equation}

\section{Controllability}

The hybrid network is controllable if the total information flow $I_i^{\rm tot}$ is constant, i.e. 
\begin{equation}
\delta I_i^{\rm cl} - \delta Q_i = 0 \, .
\label{const_flow}
\end{equation}
This seemingly simple variation equation is not very useful until $\Lambda_{j\to i}$ is specified. In practice, it depends on many protocol related factors, such as success rate from entanglement swapping, decoherence, local processing capacity, quantum memory allocation etc. If we assume that the protocol coefficients $\Lambda_{j\to i}$ are approximately fixed, i.e. $\delta \Lambda_{j\to i} \approx 0$ for all $i$, we obtain
\begin{equation}
\delta I_i^{\rm cl} + \sum_k\Lambda_{ik}\delta E_{ik} =0 \, .
\label{eq:maincontrol}
\end{equation}

In matrix notation,
\begin{equation}
\delta\bm I^{\rm cl}+\bm\Lambda\,\delta\bm e=0,
\label{eq:matrixbalance}
\end{equation}
where $\bm e$ is the vectorized upper-triangular part of $\bm E$.

\subsection{Small control variation}
\label{subsec:variation}

Let us introduce a resource-control map \cite{Sontag1998}
\begin{equation}
\bm E'=\mathcal R_{\bm u}(\bm E;\bm A),
\label{eq:Rmap}
\end{equation}
where $\bm u$ specifies the protocol choices: measurements, swapping schedules, purification operations, routing decisions, and local controls. For small control variations,
\begin{equation}
\delta \bm E = \bm B_E\,\delta\bm u + O(\norm {\delta\bm u}^2),
\label{eq:linearresource}
\end{equation}
where the matrix $\bm B_E$ is a resource-control Jacobian. 

Using Eq.~\eqref{eq:linearresource}, the variation equation Eq.~\eqref{eq:matrixbalance} becomes
\begin{equation}
\delta\bm I^{\rm cl}+\bm\Lambda\bm B_E\delta\bm u=0.
\end{equation}

A minimum-norm control solution is
\begin{equation}
\delta\bm u^\star
=
-(\bm\Lambda\bm B_E)^+
\delta\bm I^{\rm cl},
\label{eq:pseudoinverse}
\end{equation}
where $^+$ denotes the Moore--Penrose pseudoinverse \cite{Barata2012}.

\subsection{Lyapunov stability}
\label{subsec:Lyapunov}

Above assumption of small control may not be always practical, therefore we would like to consider  Eq.~\eqref{eq:matrixbalance} as a dynamical stabilization problem with respect to the constant flow. Let us define the flow error with the node-wise compensation-error vector ${\bm\epsilon}$, those components are
\begin{equation}
\epsilon_i=I_i^{\rm tot} - \delta I_i^{\rm tot}.
\end{equation}

Let us consider the Lyapunov function candidate

\begin{equation}
V(\bm\epsilon,\bm E) = \frac12\sum_i w_i\epsilon_i^2 +
\mu C_Q(\bm E) \, ,
\label{eq:Lyapunov}
\end{equation}
where $\mu$ is a positive constant which determines how strongly that resource cost contributes to the objective. 
$\boldsymbol W=\operatorname{diag}(w_1,\ldots,w_N),
\quad w_i>0$, where $W$ is the weighting matrix for the node-wise flow errors. $C_Q$ is a quantum-resource cost. 

A sufficient local stabilization condition is $\dot V\le0$. Using that $\dot{\bm\epsilon} = \dot{\bm I} + \bm\Lambda\dot{\bm e}$, we can choose a feedback law
\begin{equation}
\dot{\bm u} = -\kappa(\bm\Lambda\bm B_E)^+ \bm W\bm\epsilon \, ,
\label{eq:feedback}
\end{equation}
where $\kappa$ is the control gain.

\section{An example}

We would like to give an explicit solution of information flow stabilization in a hybrid network, those classical connectivity is described by a dense random graph with fixed adjacency matrix, i.e. $\delta A_{ij} = 0$. For such a graph, the collective term \eqref{eq:eigterm} becomes 
\begin{equation}
\bm v_1=\frac{1}{\sqrt N}\bm 1,\qquad \lambda_1=p(N-1).
\end{equation}
where $p$ is the classical edge probability, and 
\begin{equation}
g_i=\alpha v_{1,i}\simeq\frac{\alpha}{\sqrt N}\equiv g_0.
\label{eq:randomcollective}
\end{equation}
Note that $g_i$ does not depend on the edge probability due to assumption that the graph is sufficiently dense.

The collective dynamics of the classical graph \eqref{eq:randomcollective} is to be suppressed by quantum resources. Let us consider two-qubit states with constant concurrence $C_0$ \cite{Wootters1998} to be distributed over the network. For simplicity and to isolate the control mechanism we further assume no decoherence and no other practically relevant effects. 

To proceed, we need to define a bipartite incidence matrix $\bm M$ which describes connectivity of entangled links to the nodes whose information flow they can affect, i.e.
\begin{equation}
M_{i,e}=\begin{cases}1,&\text{node $i$ is incident on quantum edge $e$},\\0,&\text{otherwise}.
\end{cases}
\end{equation}
where each column represents one possible entangled link, and each row represents a network node.

For example, for a network of four nodes with quantum links $(1,3),(1,4),(2,3),(2,4)$, the matrix $\bm M$ is
\begin{equation}
\bm M=\begin{pmatrix}1&1&0&0\\0&0&1&1\\1&0&1&0\\0&1&0&1\end{pmatrix}.
\label{eq:B4}
\end{equation}

Now we can finally define the quantum information contribution as $\bm E=\gamma C_0\bm M\bm e$.

\subsection{Resource-control Jacobian}

With the network's specifications at hand, let us now return to Sec.~\ref{subsec:variation} and obtain the resource-control Jacobian explicitly. Let us take the control vector $\bm u$ to be the vector of infinitesimal entanglement link occupation changes, $u_e=\delta e$. The Jacobian is therefore
\begin{equation}
\bm B_e=\frac{\partial\bm E}{\partial\bm u}=\gamma C_0\bm M.
\label{eq:Beexplicit}
\end{equation}
Here, each column corresponds to one entangled pair and changes the quantum-flow at exactly its two endpoints.

The requirement that the quantum control eliminates classical collective dynamics leads to the compensation  equation $\bm g+\bm B_e\bm u=0$. Choosing equal control for all quantum links, $u_e=u_0$, and noticing that every node is incident on $N/2$ bipartite quantum links, we obtain
\begin{equation}
\gamma C_0\frac N2u_0=-\frac{\alpha}{\sqrt N},
\end{equation}
and therefore
\begin{equation}
u_0=-\frac{2\alpha}{\gamma C_0N^{3/2}}
\label{eq:u0}
\end{equation}
The negative sign denotes reduction of an initially available resource allocation. 

For an arbitrary connected realization with collective field $\bm g=\alpha\bm v_1(\bm A)$, the minimum-norm solution is
\begin{equation}
\bm u^\star=-\frac{\alpha}{\gamma C_0}\bm M^+\bm v_1(\bm A).
\label{eq:generalrandomsolution}
\end{equation}

\subsection{Lyapunov Analysis}

Let us now consider the same example from perspective of Lyapunov stability discussed in Sec.~\ref{subsec:Lyapunov}. 
Let us define the nodewise compensation error
\begin{equation}
\epsilon_i(t)=g_i+\gamma C_0\sum_jM_{ij}e_j(t),
\label{eq:epsilon}
\end{equation}
so that $\bm\epsilon=\bm g+\bm B_e\bm e$ and the feedback law
\begin{equation}
\dot{\bm e}=-\kappa\bm B_e^{\mathsf T}\bm\epsilon,\qquad\kappa>0.
\label{eq:lyapcontroller}
\end{equation}
Because $C_0$ is fixed and there is no memory loss or decoherence, $\bm B_e$ is constant and
\begin{equation}
\dot{\bm\epsilon}=-\kappa\bm B_e\bm B_e^{\mathsf T}\bm\epsilon.
\label{eq:errorevolution}
\end{equation}

Let us define Lyapunov function $V(\bm\epsilon)=\frac12\bm\epsilon^{\mathsf T}\bm\epsilon.
\label{eq:Vexplicit}$
Then
\begin{align}
\dot V=-\kappa\bm\epsilon^{\mathsf T}\bm B_e\bm B_e^{\mathsf T}\bm\epsilon
=-\kappa\|\bm B_e^{\mathsf T}\bm\epsilon\|^2\le0.
\label{eq:Vdot}
\end{align}
Thus the compensated state is Lyapunov stable. If $\bm B_e$ has nonzero minimum singular value $\sigma_{\min}$, then
\begin{equation}
\dot V\le-2\kappa\sigma_{\min}^2V,
\end{equation}
and
\begin{equation}
V(t)\le V(0)e^{-2\kappa\sigma_{\min}^2t},\quad
\|\bm\epsilon(t)\|\le\|\bm\epsilon(0)\|e^{-\kappa\sigma_{\min}^2t}.
\label{eq:Vexp}
\end{equation}

For the balanced complete bipartite quantum resource layer, the incidence matrix $\bm M\bm M^{\mathsf T}$ eigenvalues are $\{2n,n,\ldots,n,0\}$. The zero eigenvalue is an allocation redundancy, not an instability. The collective uniform mode $\bm1$ has eigenvalue $2n=N$ and hence the eigenvalue associated with the collective uniform mode of the matrix
\begin{equation}
\lambda_{\rm coll}(\bm B_e\bm B_e^{\mathsf T})=\gamma^2C_0^2N.
\end{equation}
This eigenvalue determines the strength of the collective direction in the resource-control coupling matrix, therefore
\begin{equation}
\epsilon_{\rm coll}(t)=\epsilon_{\rm coll}(0)e^{-\kappa\gamma^2C_0^2Nt}
\label{eq:collectivedecay}
\end{equation}
with the uniform collective error decay rate
$r=\kappa\gamma^2C_0^2N$.

\subsection{Quantum Resource Bound}

The two approaches for hybrid network controllability discussed above allow us to formulate the minimum entanglement requirement for the control procedure to be executed. Eq.~\eqref{eq:u0} suggests that the minimal concurrence is bounded, i.e.
\begin{equation}
C_0\ge C_{\min}^{(e)}
=\frac{2|\alpha|}{\gamma e_0N^{3/2}},
\end{equation}

On the other hand, a dynamical threshold follows by prescribing a minimum desired collective decay rate $r_*$ in Eq.~\eqref{eq:collectivedecay}, i.e. $\kappa\gamma^2C_0^2N\ge r_*$. Therefore
\begin{equation}
C_0\ge C_{\min}^{(r)}
=\sqrt{\frac{r_*}{\kappa\gamma^2N}},
\label{eq:Cminrate}
\end{equation}
Combining this two restrictions we obtain an overall concurrence bound
\begin{equation}
C_0\ge C_{\min}
=\max\!\left\{
\frac{2|\alpha|}{\gamma e_0N^{3/2}}, \,
\sqrt{\frac{r_*}{\kappa\gamma^2N}}
\right\}.
\label{eq:Fminfinal}
\end{equation}
It is quite interesting that the two approaches predict different scaling of the minimal concurrence with respect to the network size, i.e. $\propto(N)^{-3/2}$ and $\propto(N)^{-1/2}$.

\section{Conclusion}

Despite being generally formulated, developed mathematical framework for controlling information flow in hybrid quantum networks by redistributing entanglement resources has limitations. An e-link is not an equivalent substitution to a classical information channel, rendering it useless during network fragmentation. Certain dynamical processes that involve loss of classical connectivity $\delta A_{ij} \neq 0$ may cause correlated changes of $\Lambda$ in Eq.~\eqref{eq:quantum} introducing serious obstacles to establishing a control protocol. Also, if nodes critical for classical information flow are also critical for LOCC operations, the hybrid network becomes very sensitive to any changes in such hybrid super-hubs. In contrast, ensuring zero overlap between classically and quantum critical nodes should make hybrid quantum network dynamically robust and resilient. But designing control strategies for such complex dynamical scenario requires a comprehensive review of definitions and methods used in classical network science \cite{Siomau2025}. 

\begin{acknowledgments}
I would like to thank Eugene Demler for the opportunity to join his group at ETH Zurich, where the majority of this work was done. 
\end{acknowledgments}

\end{document}